\input{aipcheck}

\documentclass[
    ,final            
  ]
  {aipproc}

\layoutstyle{6x9}

\usepackage{epsfig}
\usepackage{units}
\usepackage[normalem]{ulem} 
\usepackage{color}
\usepackage{amsmath}
\usepackage{amssymb}

\newcommand{\bq}{\begin{equation}}
\newcommand{\eq}{\end{equation}}

\newcommand{\bqq}{\begin{eqnarray}}
\newcommand{\eqq}{\end{eqnarray}}

\newcommand{\cmsNN}{\sqrt{s_{\rm NN}}}

\newcommand{\expval}[1]{\langle #1 \rangle}
\newcommand{\etain}[1]{$|\eta|$~$<$~$#1$}

\newcommand{\pt}{\ensuremath{p_{\rm{T}}}}
\newcommand{\pta}{\ensuremath{p_{\rm T, assoc}}}
\newcommand{\ptt}{\ensuremath{p_{\rm T, trig}}}

\newcommand{\Dphi}{\Delta\varphi}
\newcommand{\Deta}{\Delta\eta}
\newcommand{\Ntrig}{N_{\rm trig}}
\newcommand{\Nassoc}{N_{\rm assoc}}
\newcommand{\icp}{I_{\rm CP}}
\newcommand{\raa}{R_{\rm AA}}
\newcommand{\iaa}{I_{\rm AA}}

\newcommand{\df}     {\mbox{${\rm d}$}}

\newcommand{\figref}[1]{Figure~\ref{#1}}

\newcommand{\bfigFullPage}{\begin{figure} \begin{center} \vspace{0pt}}
\newcommand{\bfig}{\begin{figure}}
\newcommand{\efig}{\end{figure}}
\newcommand{\btab}[1][t!]{\begin{table*}[#1] \begin{center}}
\newcommand{\etab}{\end{center} \end{table*}}

\begin{document}

\title{In-Medium Energy Loss and Correlations in Pb--Pb Collisions at $\cmsNN = \unit[2.76]{TeV}$}

\classification{25.75.Bh, 25.75.Gz, 25.75.Ld}
\keywords      {LHC, ALICE, heavy-ion collisions, jet quenching, nuclear modification factor, RAA, collective flow, Fourier decomposition, factorization relation, IAA}

\author{Jan Fiete Grosse-Oetringhaus for the ALICE collaboration}{
  address={Jan.Fiete.Grosse-Oetringhaus@cern.ch}
}

\begin{abstract}
ALICE (A Large Ion Collider Experiment) is the dedicated heavy-ion experiment at the LHC. In fall 2010, Pb--Pb collisions were recorded at a center-of-mass energy of \unit[2.76]{TeV} per nucleon pair, about 14 times higher than the energy achieved in A--A collisions at RHIC. The study of the produced hot and dense matter with an unprecedented energy density allows the characterization of the quark-gluon plasma, the deconfined state of quarks and gluons, predicted by QCD.
The study of in-medium partonic energy loss allows insights into the density of the medium and the energy-loss mechanisms. This paper presents results based on inclusive spectra as well as two and more-particle correlations of charged particles. These are well suited to assess in-medium effects, ranging from the suppression of particles ($\raa$) and away-side jets ($\iaa$) at high $\pt$ to long-range phenomena attributed to collective effects like the ridge at low $\pt$.
The analysis is discussed and the results are presented in the context of earlier RHIC measurements where appropriate.
\end{abstract}

\maketitle
%
%
%
The objective of the study of ultra-relativistic heavy ion-collisions  is the characterization of
the quark--gluon plasma, the deconfined state of quarks and gluons. 
This paper discusses recent measurements by ALICE quantifying the modification of particle yields by the medium.
The suppression of charged hadrons
expressed as the nuclear modification factor $R_{\rm AA}$ 
is presented.
The correlation of azimuthally back-to-back dihadrons enables a systematically different study of energy loss. Due to the trigger condition, the path length in the medium of the recoiling parton is biased to be longer than the average for the inclusive measurements.
Triggered correlations are studied to assess at which $\pt$ collective effects dominate
and in which region jet-like correlations contribute most. In the jet-dominated
regime, at higher $\pt$, near- and away-side yields are measured and compared to a pp reference ($\iaa$).

\subsection{Detector and Data Sample}

The Inner
Tracking System (ITS) and the Time Projection Chamber (TPC) of the ALICE detector \cite{alice} are used for vertex finding and
tracking. Forward scintillators (VZERO) determine the centrality of the collisions.
The main tracking information is recorded by the TPC which has a uniform acceptance in azimuthal angle ($\phi$) and allows good-quality track reconstruction within \etain{1.0}. 
For the $\raa$-analysis, the information from the ITS is used in the tracking procedure as well. This results in excellent primary vertex and $\pt$ resolution and thus secondary rejection.
The $\pt$ resolution can be parametrized by $\sigma_{\pt} / \pt \approx 0.002 \cdot \pt$, e.g., about 10\% at $\pt = \unit[50]{GeV/\emph{c}}$, and was verified with high momentum cosmic-muon tracks and invariant mass distributions of $K^0_s$ and $\Lambda$.

For the two-particle correlation analysis, a different strategy is chosen to achieve a uniform acceptance. Tracks reconstructed with only the TPC information are used. Their $\pt$ is constrained with the reconstructed vertex information.

ALICE recorded about 30 million minimum-bias Pb--Pb events at $\cmsNN = \unit[2.76]{TeV}$ in fall 2010 as well as a reference sample of 70 million minimum-bias pp events at the same energy per nucleon pair in March 2011.

\subsection{Nuclear modification factor $\raa$}

\bfig
	\includegraphics[width=0.48\linewidth]{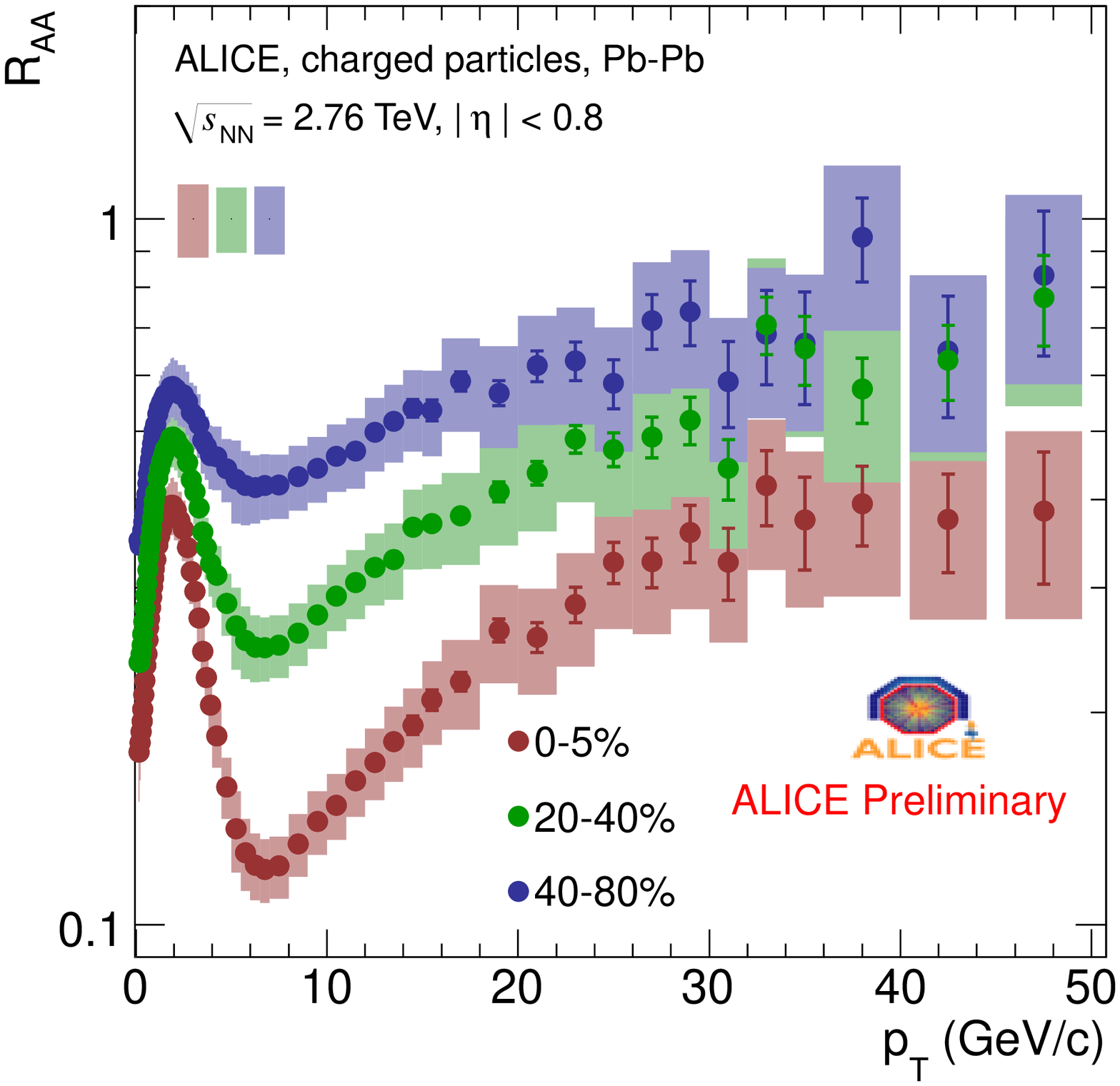}
	\hfill
	\includegraphics[width=0.48\linewidth]{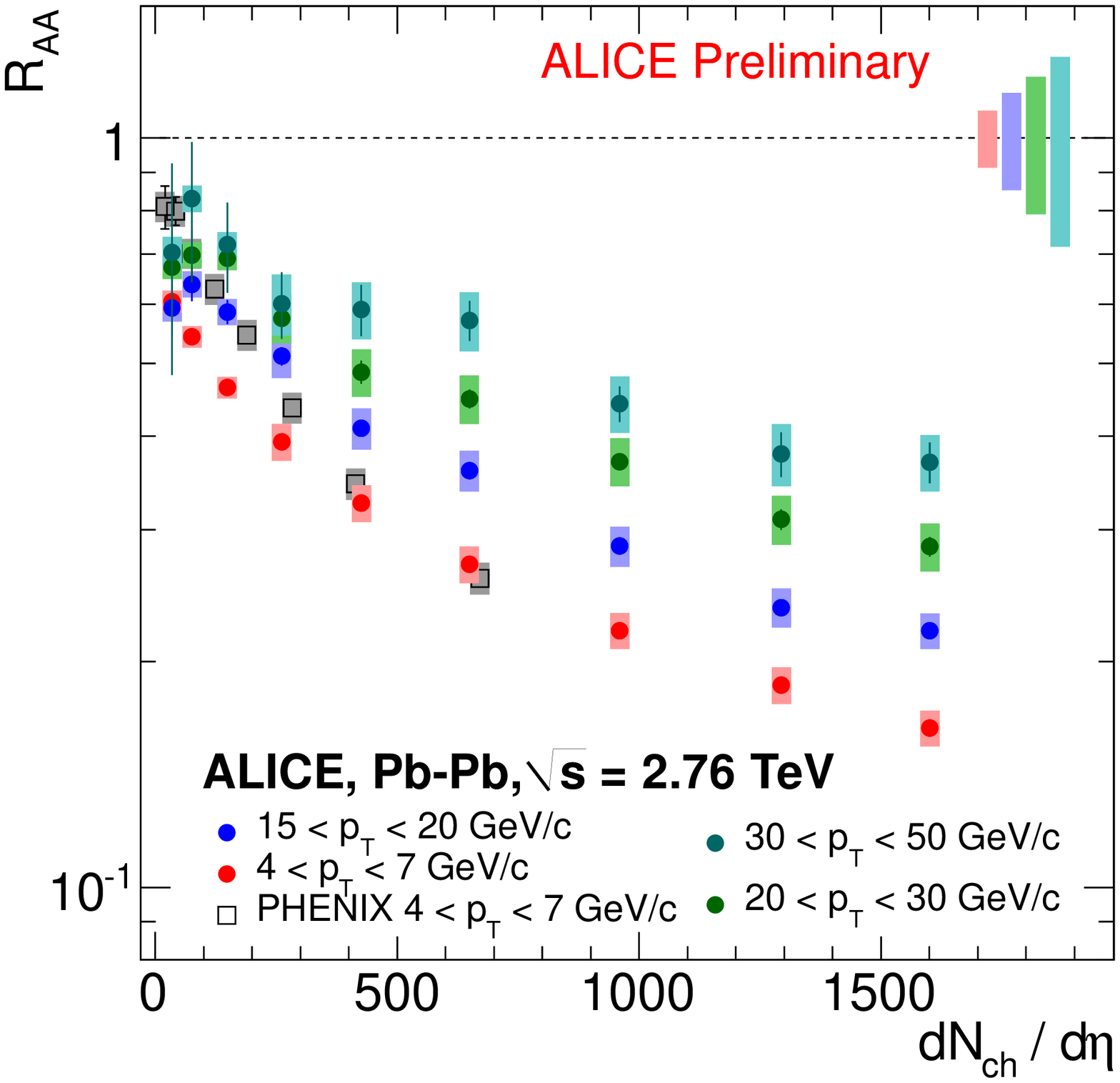}
	\caption{\label{raa} Left panel: $\raa$ for different centrality intervals as function of $\pt$ of Pb--Pb collisions at $\cmsNN = \unit[2.76]{TeV}$. Right panel: $\raa$ for different $\pt$-intervals as function of the charged-particle pseudorapidity density $dN_{\rm ch}/d\eta$ compared to results from RHIC. Shaded areas denote systematic uncertainties.}
\efig

The nuclear modification factor is defined as
\bq
	\raa = \frac{dN_{\rm Pb-Pb}/d\pt}{\langle N_{\rm coll} \rangle dN_{\rm pp}/d\pt}
\eq
where $dN_{\rm Pb-Pb (pp)}/d\pt$ is the inclusive charged particle yield for Pb--Pb (pp) collisions as function of transverse momentum $\pt$, averaged over a percentile bin in centrality. $\langle N_{\rm coll} \rangle$ is the average number of binary nucleon--nucleon collisions in the centrality interval and is determined with a Monte-Carlo Glauber model implementation \cite{glauber}.
The Pb--Pb distribution is measured up to $\pt = \unit[50]{GeV/\emph{c}}$. Due to the available statistics the pp reference distribution is only available up to about \unit[40]{GeV/\emph{c}} and is extrapolated by a modified Hagedorn function \cite{hagedorn}. 

\figref{raa} (left panel) shows $\raa$ for different centrality intervals. A minimum is observed at $\pt = \unit[6-8]{GeV/\emph{c}}$ in all of them, above which $\raa$ rises monotonically. The observed suppression in central collisions is significantly larger than that at RHIC \cite{raarhic1, raarhic2} (plot not shown, see \cite{harryqmproc}). 
It is interesting to evaluate $\raa$ in $\pt$-intervals and study it as a function of the charged-particle pseudorapidity density as shown in the right panel of \figref{raa}. The data are overlaid with similar measurements from PHENIX in the region of the strongest suppression ($4 < \pt < \unit[7]{GeV/\emph{c}}$) which shows that at the same $dN_{\rm ch}/d\eta$ the suppression is comparable. 
For more details see \cite{harryqmproc,jacekqmproc}.

\subsection{Di-Hadron Correlations}
In triggered azimuthal correlations, a particle from one $\pt$ region is chosen and called \emph{trigger particle}. So-called \emph{associated particles} from another $\pt$ region are correlated to the trigger particle, where $\pta < \ptt$. With these definitions one calculates the per-trigger yield
\bq
  Y(\Dphi) = \frac{1}{\Ntrig} \frac{\df \Nassoc}{\df \Dphi} \label{phi_yield}
\eq
and the correlation function
\bq
  C(\Deta, \Dphi) = \left( \frac{1}{N_{\rm pairs}} \frac{d\Nassoc}{d\Dphi d\Deta} \right)_{\rm same} / \left( \frac{1}{N_{\rm pairs}} \frac{d\Nassoc}{d\Dphi d\Deta} \right)_{\rm mixed}.
\eq
$\Ntrig$ is the number of trigger particles to which $\Nassoc$ particles are associated at $\Dphi = \phi_{\rm trig} - \phi_{\rm assoc}$ and $\Deta = \eta_{\rm trig} - \eta_{\rm assoc}$. $N_{\rm pairs}$ is the total number of pairs. The subscript same (mixed) indicates that pairs are formed from the same (different) events.

\bfig
  \includegraphics[width=\linewidth,trim=0 0 0 12,clip=true]{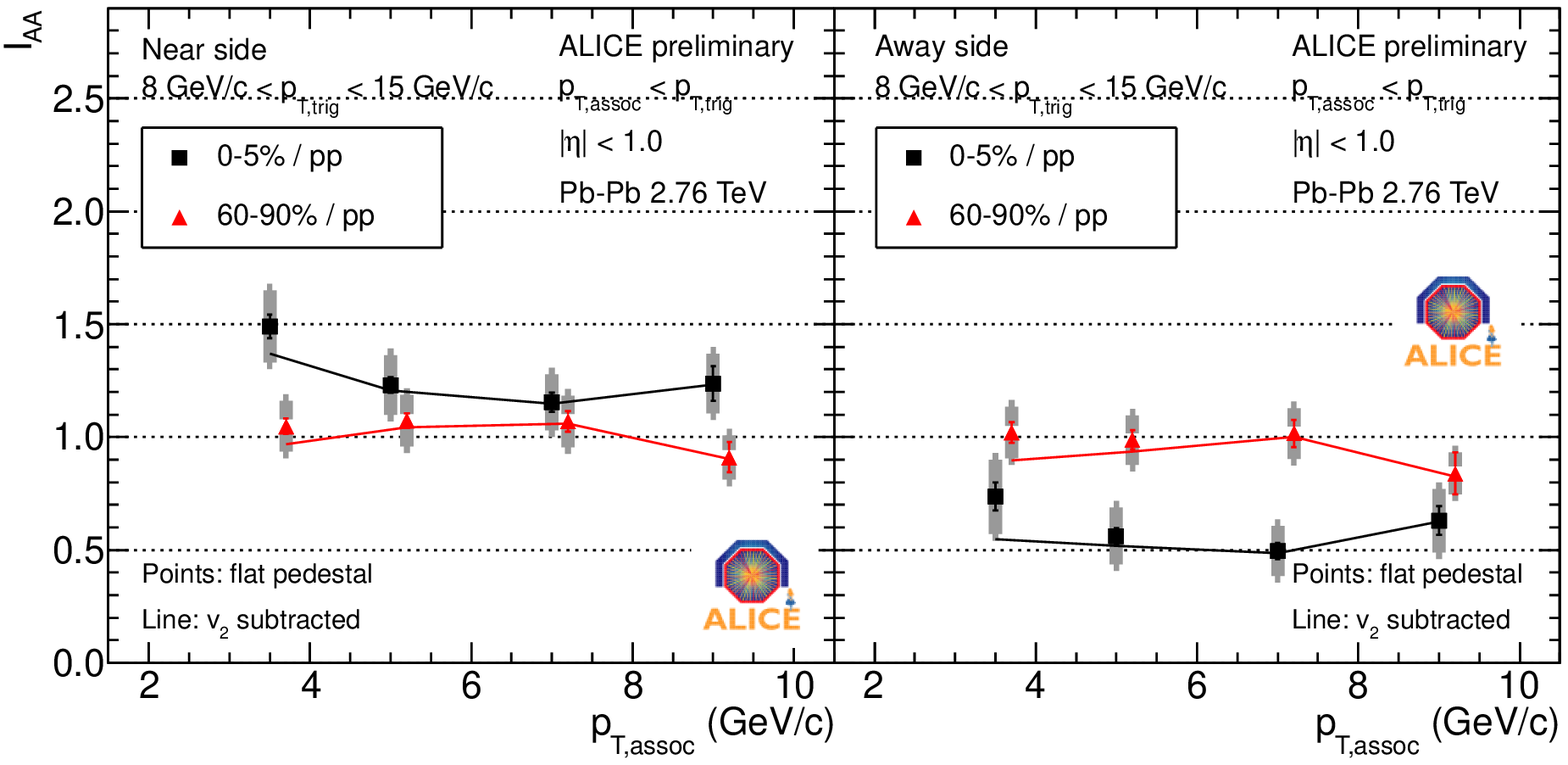}
  \caption{\label{fig_iaa} $\iaa$: the data points are calculated with a uniform pedestal; the line is based on $v_2$ subtracted yields. Shaded areas denote systematic uncertainties.}
\efig

\paragraph{Correlated yields}

\bfig
  \includegraphics[width=\linewidth,trim=0 0 0 12,clip=true]{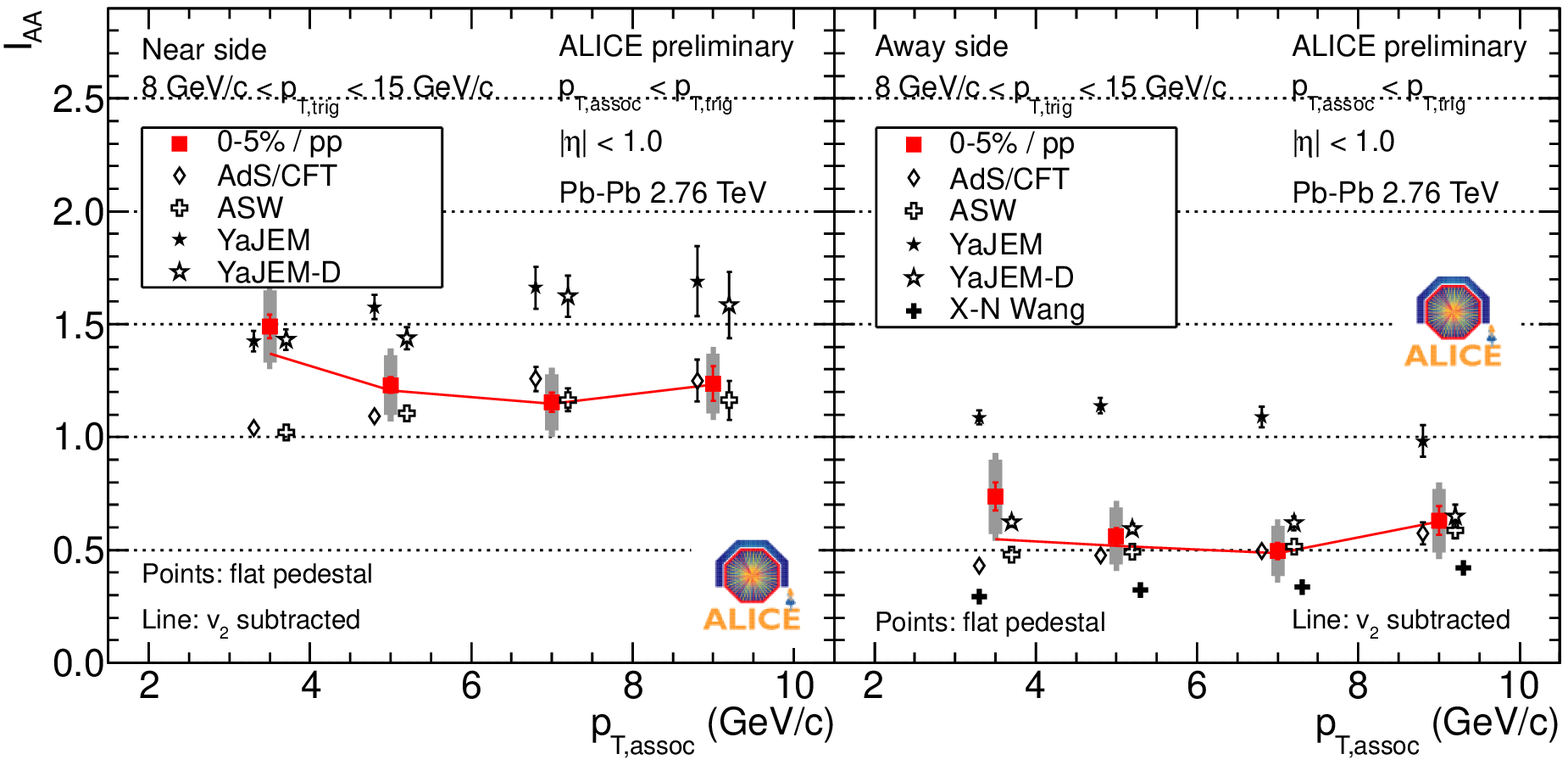}
  \caption{\label{fig_iaatheory} $\iaa$ compared to theoretical calculations.}
\efig

At sufficiently high $\pt$ that collective effects are expected to be small and jet-like correlations dominate ($\ptt > \unit[8]{GeV/\emph{c}}$; $\pta > \unit[3]{GeV/\emph{c}}$), the effect of the medium on the jet fragmentation can be studied.
This is assessed by measuring the associated yield to a trigger particle and by calculating yield ratios between Pb--Pb and pp collisions. To remove uncorrelated background from the associated yield, the pedestal value needs to be determined.
This is done by fitting the region close to the minimum of the $\Dphi$ distribution ($\Dphi \approx \pm \frac{\pi}{2}$) with a constant and using this value as pedestal (zero yield at minimum -- ZYAM). 
We estimate and subtract the $v_2$ contribution to the signal using an independent measurement \cite{newflow}.
Subsequent to the pedestal (and optionally $v_2$) subtraction, the near- and away-side yields are integrated within $\Dphi$ of $\pm 0.7$ and $\pi \pm 0.7$, respectively. See Fig.~7 in \cite{jfgoqmproc} for an illustration.

To quantify the effect of in-medium partonic energy loss, the associated-yield ratio ($\iaa$) in Pb--Pb and pp collisions is calculated.
\figref{fig_iaa} shows $\iaa$ for central and peripheral collisions using the uniform pedestal (data points) and $v_2$ subtracted yields (lines).
The only significant difference is in the lowest bin of $\pta$ while all other bins show a negligible difference. This confirms the small bias due
to flow anisotropies in this $\pt$ region.
Note that only $v_2$ is considered here, although contributions from other harmonics may be significant, particularly for central events. In central collisions, away-side suppression from in-medium energy loss is seen ($\iaa \approx 0.6$), as expected. On the near-side, there is an enhancement above unity ($\iaa \approx 1.2$) that was not observed with significance at lower collision energies \cite{iaastar}. In peripheral collisions, both near- and away-side $\iaa$ are consistent with unity.

In addition, the ratio of yields in central and peripheral collision, $\icp$, has been calculated (plot not shown, see \cite{jfgoqmproc}) and is found to be consistent with $\iaa$ in central collisions, exhibiting both the near-side enhancement and the away-side suppression.

\figref{fig_iaatheory} shows $\iaa$ compared to theoretical calculations for different energy loss scenarios \cite{renkprediction,wangprediction}. The near-side enhancement is reproduced by energy loss modeled in AdS/CFT (cubic path length dependence) and ASW (quadratic dependence). YaJEM (linear dependence) as well as YaJEM-D yield too large values. The away-side suppression is reproduced by AdS/CFT, ASW and YaJEM-D; YaJEM is too high and the calculation from X.~N.~Wang yields slightly too low values.

For more details about this measurement see \cite{jfgoqmproc}.

\paragraph{Disentangling collective and jet fragmentation effects}

\bfig
  \includegraphics[width=0.48\linewidth,trim=0 0 0 10,clip=true]{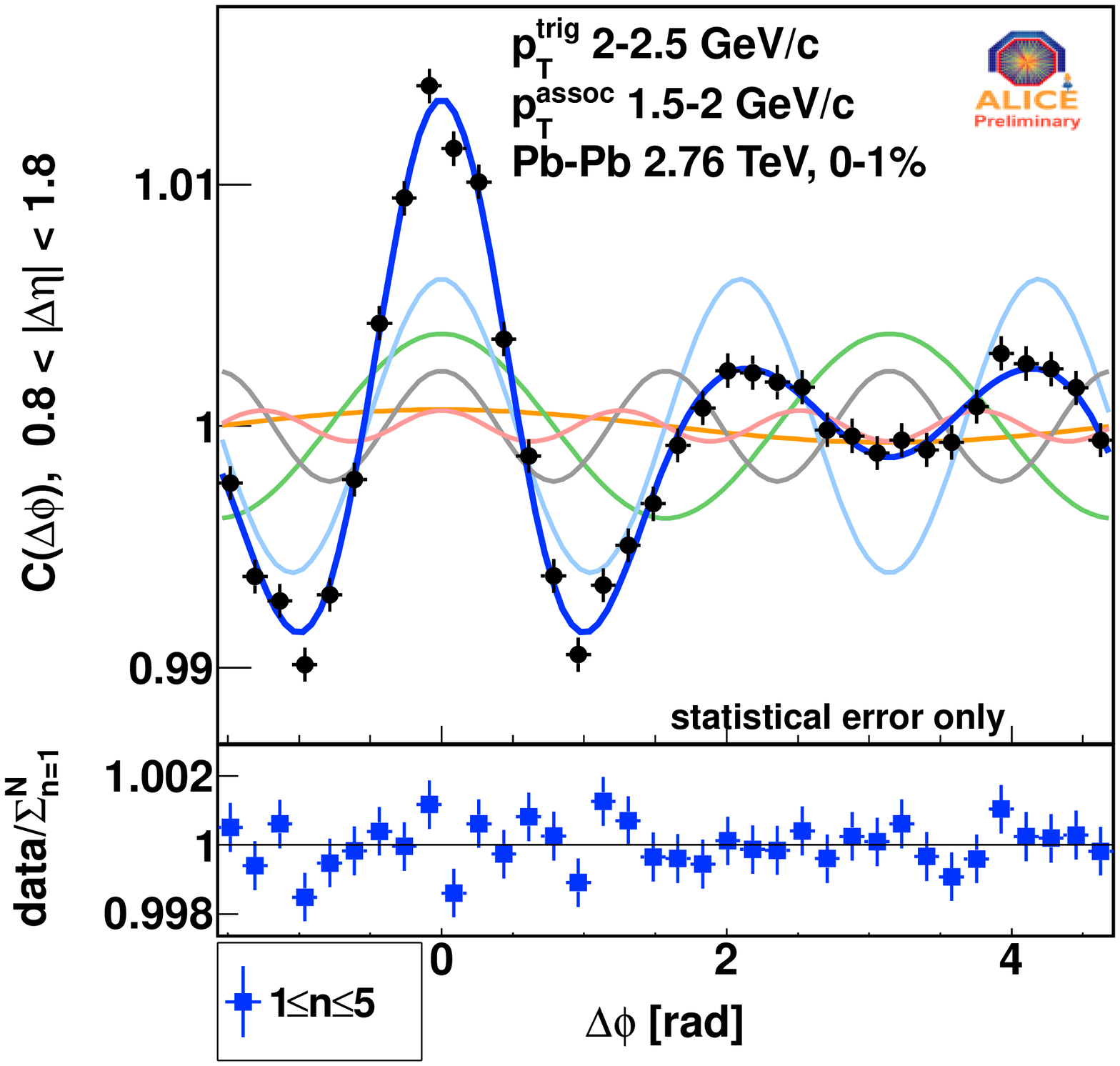}
  \hfill
  \includegraphics[width=0.48\linewidth,trim=0 0 0 10,clip=true]{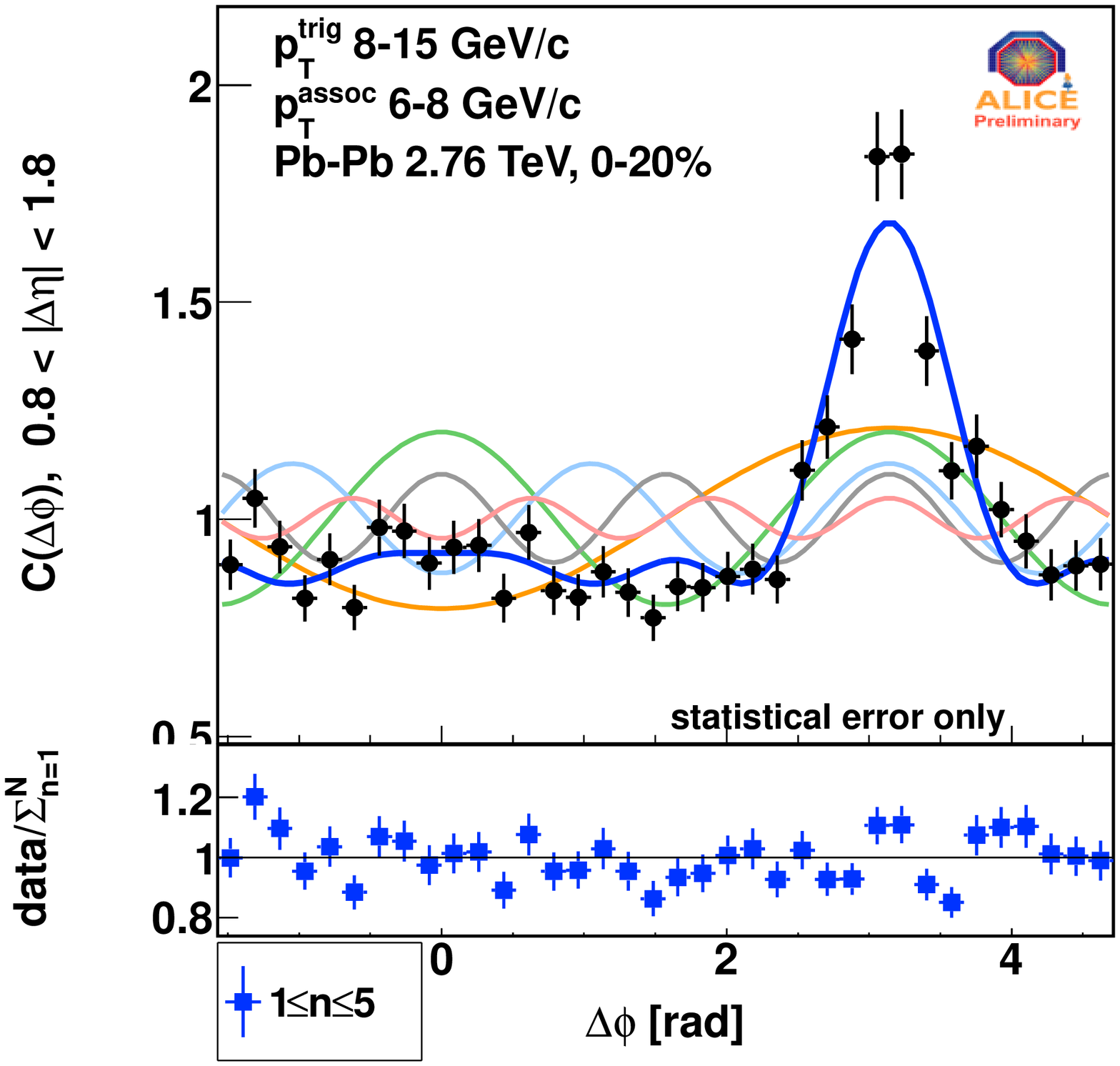}
  \hspace{0.03\linewidth}
  \caption{\label{fourier_examples} Correlation function (points) and its Fourier decomposition (lines) in $0.8 < |\Deta| < 1.8$ for a low $\pt$ bin in \unit[0-1]{\%} centrality (left) and a high $\pt$ bin (left) in \unit[0-20]{\%} centrality (right). The $\pt$ ranges are indicated on the plot.
  The first five Fourier harmonics are shown as well as their sum. The lower part shows the residuals between the  data and the sum of the harmonics.}
\efig

The long-range correlation region within $0.8 < |\Deta| < 1.8$ in Pb--Pb collisions is projected onto $\Dphi$ and decomposed into Fourier coefficients using the relation:
\bq
  V_{n\Delta} = \expval{\cos n \Dphi} = \frac{\int d\Dphi C(\Dphi) \cos n \Dphi}{\int d\Dphi C(\Dphi)}.
\eq
\figref{fourier_examples} shows $C(\Dphi)$ in a low $\pt$ and a high $\pt$ bin, for different centrality bins. The first five Fourier components are shown as well as their sum which describes the correlation well at low $\pt$ and reasonably at high $\pt$. At low $\pt$, no significant improvement is observed by including higher coefficients. In the low $\pt$ bin (0-1\% centrality) a double-humped structure on the away-side is observed (notably, even without background subtraction) which is discussed in more detail in \cite{newflow, decompositionpaper}. The extracted coefficients (not shown, see \cite{decompositionpaper}) are a compact description of the measured data. They increase with increasing $\pt$ and then decrease for $\ptt > \unit[5]{GeV/\emph{c}}$. The odd terms become negative at large $\pt$, which can be attributed to the influence of the away-side jet. From central to peripheral collisions, $V_{2\Delta}$ rises most.
To assess which part of the correlation is caused by collective effects, we test the factorization relation:
  $V_{n\Delta}(\pta,\ptt) \stackrel{?}{=} v_{n}(\pta) \cdot v_{n}(\ptt)$
which is valid if the two-particle correlation is dominated by correlations with a common plane of symmetry. This is not the case for jet-like correlations where a few particles are correlated by fragmentation\footnote{Indirect correlations exist, e.g. length-dependent quenching which has the largest influence on $V_{2\Delta}$.}.
We first determine the coefficients $V_{n\Delta}(\pta,\ptt)$
as a function of $\pta$ and $\ptt$, and then carry out an independent
global fit allowing only one value per $\pt$-bin, i.e. $v_{n}(\pt)$.
\figref{fourier_factorization} presents their comparison, which tests the factorization relation for $V_{2\Delta}$ in \unit[0-10]{\%} centrality. The bottom part of the figure shows the ratio between the two indicating that the factorization works well at low to intermediate $\pt$ (up to $\unit[3-4]{GeV/\emph{c}}$ depending on centrality) and breaks down above that value where jet-like correlations are expected to dominate.
This trend is followed for $n>2$ as well, but $V_{1\Delta}$ does not seem to follow a clear factorization pattern even in flow-dominated regimes (plots not shown, see \cite{decompositionpaper}).
In the $\pt$ region where the factorization holds, the extracted coefficients are consistent with other measurement of flow coefficients, e.g. \cite{newflow}.
More details can be found in \cite{decompositionpaper}.

\bfig
    \includegraphics[width=0.6\linewidth,trim=0 0 0 5,clip=true]{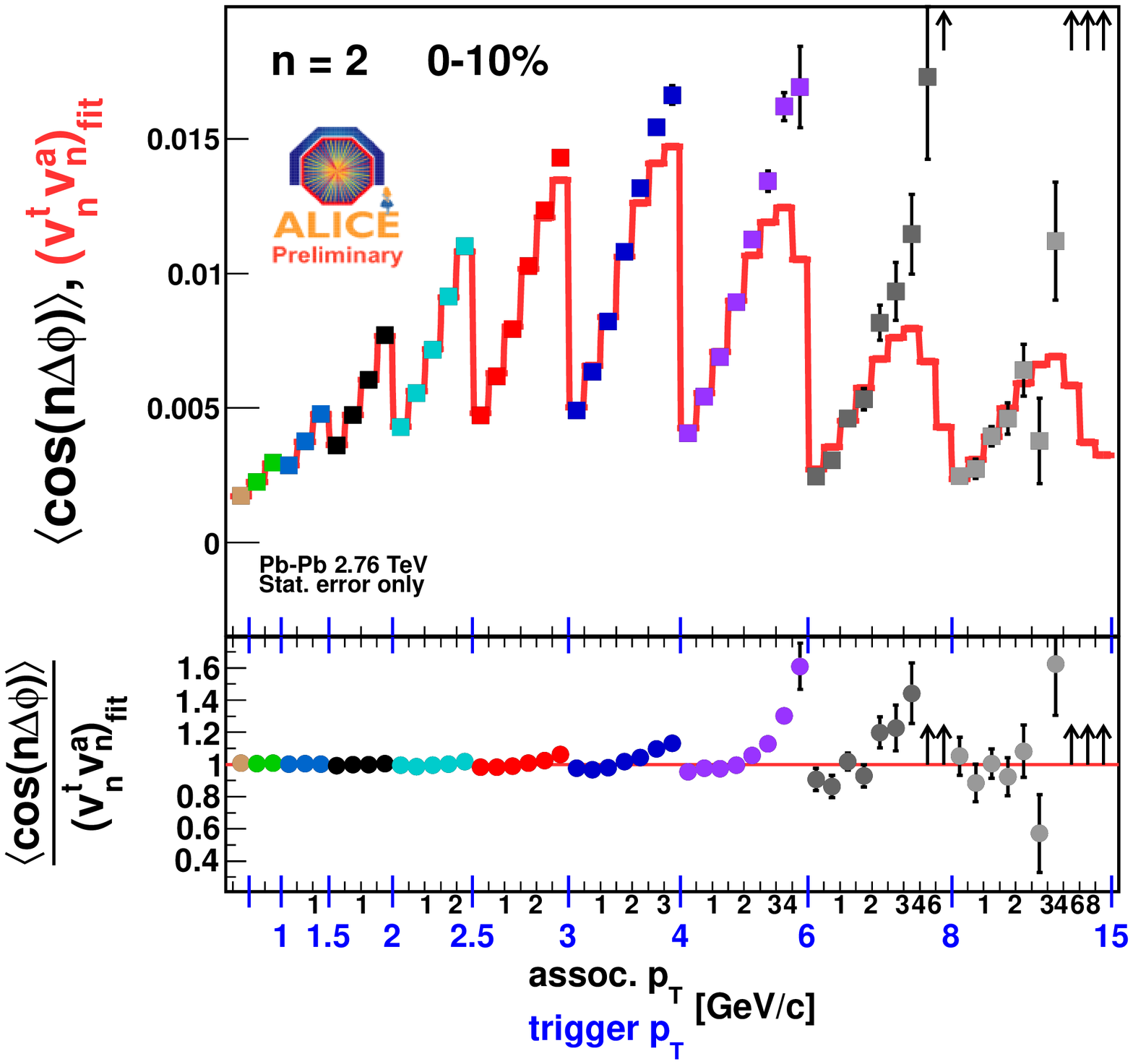}
    \caption{\label{fourier_factorization} Test of the factorization relation for: $V_{2\Delta}(\pta,\ptt)$ (points); $v_{2}(\pta) \cdot v_{2}(\ptt)$ (line). The bottom panel shows their ratio.}
\efig

\subsection{Summary}

Measurements based on inclusive production and correlations of charged hadrons that characterize the hot and dense medium created in central Pb--Pb collisions at the LHC have been presented. A strong suppression in the inclusive $\pt$ distribution is observed. Around $\pt = \unit[6-8]{GeV/\emph{c}}$ $\raa$ reaches a minimum of about 0.12. Di-hadron correlations allow to extract the modification of the particle yield associated to a high $\pt$ trigger particle ($\iaa$). On the away-side suppression is observed, $\iaa \approx 0.6$, while on the near-side a moderate enhancement ($\iaa \approx 1.2$) is measured. $V_{N\Delta}$ coefficients are extracted by a Fourier decomposition of the two-particle correlation function. Their factorization is tested which indicates that collective effects dominate for $\pt \lesssim \unit[3-4]{GeV/\emph{c}}$. Above this momentum the factorization relation breaks down and jet-like correlations seem to dominate. These measurements provide strong constraints on the energy-loss mechanism in the hot and dense matter. Further theoretical modeling and extension of the kinematical reach of these measurements are needed to discriminate between the different energy-loss scenarios.

\bibliographystyle{aipproc}

\begin{thebibliography}{19}

\bibitem{alice}   K.~Aamodt {\it et al.} [ALICE Collaboration],
  JINST {\bf 3 } (2008)  S08002.

\bibitem{glauber} A.~Toia for the ALICE collaboration, Quark Matter 2011 Proceedings.

\bibitem{hagedorn} R. Hagedorn, Riv. Nuovo Cim. \textbf{6} (1983) 1.

\bibitem{raarhic1} J. Adams et al., STAR Collaboration, Phys. Rev. Lett. \textbf{91}, 172302 (2003).
	
\bibitem{raarhic2} S.S. Adler et al., PHENIX Collaboration, Phys. Rev. C \textbf{69}, 034910 (2004).

\bibitem{harryqmproc} H.~Appelsh\"auser for the ALICE collaboration, Quark Matter 2011 Proceedings.

\bibitem{jacekqmproc} J.~Otwinowski for the ALICE collaboration, Quark Matter 2011 Proceedings.

\bibitem{newflow}
  K.~Aamodt {\it et al.} [ALICE Collaboration],
  arXiv:1105.3865 [nucl-ex].

\bibitem{jfgoqmproc} J.F.~Grosse-Oetringhaus for the ALICE collaboration, Quark Matter 2011 Proceedings.

\bibitem{iaastar}   J.~Adams {\it et al.} [STAR Collaboration],
  Phys.\ Rev.\ Lett.\  {\bf 97 } (2006)  162301.

\bibitem{renkprediction} T.~Renk and K.~J.~Eskola,
  arXiv:1106.1740 [hep-ph].

\bibitem{wangprediction} X.~N.~Wang, private communication; see also
  H.~Zhang, J.~F.~Owens, E.~Wang and X.~N.~Wang,
  Phys.\ Rev.\ Lett.\  {\bf 98} (2007) 212301.

\bibitem{decompositionpaper}   K.~Aamodt {\it et al.} [ALICE Collaboration],
  arXiv:1109.2501 [nucl-ex].



\end{thebibliography}

\end{document}